\documentclass[11pt,a4paper]{iopart}
\usepackage{cite}
\usepackage{graphicx} 
\usepackage{xcolor}

\usepackage[utf8]{inputenc}

\usepackage{iopams}

\expandafter\let\csname equation*\endcsname\relax

\expandafter\let\csname endequation*\endcsname\relax 

\usepackage{amsmath}
\usepackage{amssymb}
\usepackage{hyperref}
\usepackage{enumitem}

\makeatletter
\def\@mkboth#1#2{}
\newlength\appendixwidth
\newcommand{\patchl@section}{%
  \settowidth{\appendixwidth}{\textbf{Appendix }}%
  \addtolength{\appendixwidth}{1.5em}%
  \patchcmd{\l@section}{1.5em}{\appendixwidth}{}{\ddt}%
}
\makeatother

\begin{document}

\title[Edwards-Wilkinson interface with diffusing diffusivity]{Exact height distribution in one-dimensional Edwards-Wilkinson interface with diffusing diffusivity}
\author{David S. Dean}
\address{Univ. Bordeaux, CNRS, LOMA, UMR 5798, F-33400, Talence, France.\\
Email: david.dean@u-bordeaux.fr}

\author{Satya N. Majumdar}
\address{LPTMS, CNRS, Universit\'e Paris-Sud, Universit\'e Paris-Saclay, 91405 Orsay, France\\
Email: satyanarayan.majumdar@cnrs.fr}

\author{Sanjib Sabhapandit}
\address{Raman Research Institute, Bangalore 560080, India\\
Email: sanjib@rri.res.in}

\begin{abstract}
   We study the height distribution of a one-dimensional Edwards-Wilkinson  interface in the presence of a stochastic diffusivity $D(t)=B^2(t)$, where $B(t)$ represents a one-dimensional Brownian motion at time $t$. The height distribution at a fixed point in space is computed analytically. The typical height $h(x,t)$ at a given point in space is found to scale as $t^{3/4}$ and the distribution $G(H)$ of the scaled height $H=h/t^{3/4}$ is symmetric but with a nontrivial shape: while it approaches a nonzero constant quadratically as
   $H\to 0$, it has a non-Gaussian tail that decays exponentially for large $H$. We show that this exponential tail is rather robust and holds for a whole family of linear interface models parametrized by a dynamical exponent $z>1$, with $z=2$ corresponding to the Edwards-Wilkinson model. 
\end{abstract}

\maketitle

\section{Introduction}

One of the simplest models of a polymer chain in a solvent (without excluded volume or hydrodynamic interactions)  is the celebrated Rouse model of beads connected by harmonic springs~\cite{rouse1953theory, de1976dynamics}. Denoting the position of the $n$-th monomer/bead by $h_n(t)$ in an infinite one-dimensional chain, the energy of the chain is given by $E=(\kappa/2) \sum_n (h_n-h_{n-1})^2$, where $\kappa$ represents the strength of the harmonic interactions. The stochastic  dynamics (the so-called model-A) of this chain is given by the Langevin equation~\cite{rouse1953theory}   
 \begin{equation}
    \frac{dh_n}{dt} = \kappa\,  [h_{n+1}(t) + h_{n-1}(t) - 2 h_n(t))] + \sqrt{2D}\, \eta_n(t). 
    \label{eq:Rouse}
\end{equation}
where $\eta_n(t)$ is a Gaussian white noise with zero mean and the correlator $\langle \eta_n(t) \eta_m(t') \rangle = \delta_{n,m}\, \delta(t-t')$ and $D$ represents the diffusion constant.  Without the interaction, i.e., for $\kappa=0$, the position $h_n(t)$ of the $n$-th monomer undergoes normal diffusion with $\langle h_n^2(t)\rangle =2D t$. However, in the presence of the harmonic interaction, i.e., for $\kappa>0$, the position of a tagged monomer, e.g, the $n$-th monomer, grows anomalously slowly as $h_n(t)\sim t^{1/4}$ for $t\gg 1/\kappa$.  This can be easily seen by computing exactly the mean squared displacement (MSD) $\langle h^2_n(t) \rangle $ (assuming all monomers start initially at $h_n(0)=0$).  Moreover, due to the linearity of \eref{eq:Rouse},  the full distribution of $h_n(t)$ is Gaussian at all times with this variance  $\langle h^2_n(t) \rangle $.

While the short-time behavior ($t\ll 1/\kappa$) of the variance $\langle h^2_n(t) \rangle $ is affected by the discrete nature of the beads in \eref{eq:Rouse}, the late-time ($t\gg 1/\kappa$) 
behavior of the variance $\langle h^2_n(t) \rangle \sim t^{1/2}$ can be easily obtained from a continuum coarse-grained version of \eref{eq:Rouse}, where the label $n$ of a monomer is replaced by a continuous space index $x$ with the identification $h_n(t)\to h(x,t)$.  Replacing the discrete Laplacian by its continuous counterpart, \eref{eq:Rouse} reduces to the effective hydrodynamic equation 
\begin{equation}
    \frac{\partial h}{\partial t} =\Gamma  \frac{\partial^2 h}{\partial x^2}  + \sqrt{2 D}\,\eta(x,t)\,,
    \label{eq:EW}
\end{equation}
where $\Gamma \propto \kappa$ and  $\eta(x,t)$ is a Gaussian white noise with zero mean and the correlator $\langle \eta(x,t)\eta(x',t')\rangle =\delta(x-x')\delta(t-t')$ with $D$ representing the amplitude of the noise.

This hydrodynamic limit of the Rouse dynamics in \eref{eq:EW} corresponds to another celebrated model of a fluctuating interface in one dimension, known as the Edwards-Wilkinson (EW) model~\cite{edwards1982surface}, where $h(x,t)$ represents the height of the interface at position $x$ at time $t$ on an infinitely long substrate.
This interface model has also been studied for many decades with many applications~\cite{barabasi1995fractal, krug1997origins, halpin1995kinetic}.  In the EW model, due to the linearity of~\eref{eq:EW}, the full height distribution at fixed point $x$ is given at all times $t$ by the simple Gaussian 
\begin{equation}
    p(h,t)= \frac{1}{\sqrt{2 \pi V(t)}}\, \exp\left( - \frac{h^2}{2 V(t)}\right),
    \label{dist-H}
\end{equation}
where the variance $V(t)$ can be easily computed at all times. In an infinite system,  starting from a flat initial condition $h(x,0)=0$, by taking Fourier transform of \eref{eq:EW}, one can easily show  that the variance $V(t)$ grows algebraically \emph{at all times} $t$ as
\begin{equation}
V(t)=\langle h^2(x,t)\rangle =\sqrt{\frac{2}{\pi\, \Gamma}} \,D \,t^{2\beta}\quad\text{where}~~\beta = \frac{1}{4}\,.
\label{eq:Vt}
\end{equation}
While this result holds at all times $t$ for the continuous EW model in an infinite system, it describes the MSD of a tagged monomer in the Rouse model only at late times when $t\gg 1/\kappa$.
Thus, to summarise,  the position distribution of a tagged monomer in the Rouse model is Gaussian at all times, even though the MSD grows subdiffusively as $t^{1/2}$ at late times ($t\gg 1/\kappa$).

In this simple Rouse chain/EW model in one dimension, the noise amplitude, i.e., the diffusivity $D$ in~\eref{eq:EW} is taken to be a constant. However, in many dynamically heterogeneous systems, the diffusivity may also change stochastically with time. Examples include transport in complex environments such as  the motion of polystyrene beads on the surface of a lipid bilayer tube~\cite{WABG09},  liposomes moving in a nematic solution of actin filaments~\cite{WKBG012},  diffusion of tracer molecules on polymer thin films~\cite{BSSDSNC2013}, 
transport of a tracer in hard-sphere colloidal suspensions~\cite{guan2014even}, dynamics of nanoparticles~\cite{he2013diffusive},  many others. In simpler systems, the diffusivity can vary due to proximity with hard walls~\cite{czajka2019effects, alexandre_2023}, as well as due to changes in the conformation or orientation of the diffusing particle~\cite{han2006brownian, Munk2009}. Such stochasticity in diffusivity has been studied theoretically in several single particle models that exhibit `Brownian yet non-Gaussian' behavior~\cite{CS14}, with diverse applications ranging from intracellular transport all the way to finance--for other applications see the review~\cite{SGMSS2020}.  For example, setting $\kappa=0$ in the Rouse model~\eref{eq:Rouse}  and dropping the index $n$ of $h_n(t)$, the position of any single monomer evolves as
\begin{equation}
    \frac{dh}{dt} = \sqrt{2 D(t)} \, \xi (t)\,, 
    \label{eq:xt}
\end{equation}
where $\xi(t)$ is a Gaussian white noise with zero mean and correlator $\langle\xi(t)\xi(t')\rangle=\delta(t-t')$ and $D(t)$ represents a stochastic process that remains positive at all times. One of the main motivations for such `diffusing diffusivity' models is as follows. Several experimental systems involving a single particle, such as a colloid in a non-Newtonian fluid, while the mean squared displacement (MSD) of the position grows as normal diffusion, i.e., $\langle h^2(t)\rangle \sim  t$ at late times, the position distribution often exhibits non-Gaussian tails~\cite{WABG09, WKBG012}. The `diffusing diffusivity' models have been put forward to explain,  such `Brownian yet anomalous' diffusion~\cite{CS14} in a simple setting. In the literature, several choices of $D(t)$ have been studied analytically~\cite{SGMSS2020}. When $D(t)$ represents a positive stochastic process that does not grow with time (such as the square of an Ornstein-Uhlenbeck process), it has been shown that while the MSD $\langle h^2(t)\rangle \sim t$ at late times, the scaled position variable $h/\sqrt{t}$ has a distribution with a non-Gaussian tail, thus explaining the `Brownian yet non-Gaussian' behavior~\cite{chechkin2017brownian, tyagi2017non, lanoiselee2018model, SGMSS2020, barkai2020packets, pacheco2021large, hamdi2024laplace, singh2024emergence, gueneau2025large, hidalgo2021cusp,sposini2018first,jain2017diffusing,jain2016diffusingsurvival,yin2021non }. 

Other choices of $D(t)$ that are positive but that typically grow with time have also been studied in the literature in this single particle setting. In this case, the MSD may not grow linearly with time as in normal diffusion, but the position distribution is still anomalous with a non-Gaussian tail. A simple choice of such growing $D(t)$ that is fully solvable analytically corresponds to  
 choosing $D(t)$  as the square of a Brownian motion~\cite{SGMSS2020}  , i.e., 
\begin{equation}
    D(t) = B^2(t),
    \label{eq:Dt}
\end{equation}
where $B(t)$ is a standard Brownian motion in one dimension evolving as
\begin{equation}
    \frac{dB}{dt} = \sqrt{2}\,\chi(t),
    \label{eq:bm}
\end{equation}
and $\chi(t)$ is a Gaussian white noise with zero mean and two-time correlation function $\langle  \chi(t) \chi(t')\rangle =\delta(t-t')$. We assume that $B(0)=0$. For a given history of $D(t)$, the position $h(t)$ in \eref{eq:xt} has a Gaussian distribution 
\begin{equation}
    P(h,t|\{D(\tau)\})= \frac{1}{\sqrt{2\pi V_0(t)}} e^{-h^2/(2 V_0(t))}~,
    \label{eq:pxt|d}
\end{equation}
where the variance $V_0(t)$ is given by the Brownian functional 
\begin{equation}
    V_0(t) = 2 \int_0^t B^2(\tau)\, d\tau.
\end{equation}
Clearly $V_0(t)$ scales as $t^2$, since $B(\tau)\sim \sqrt{\tau}$. The distribution of $V_0(t)$ can then be expressed in a scaling form
$P(V_0, t) = t^{-2}\, F_0(V_0/t^2)$, where the function $F_0(z)$ can be computed explicitly using the backward Feynman-Kac formula. Hence, the position distribution $P(h,t)$ at time $t$, using \eref{eq:pxt|d},  can be expressed as
\begin{equation}
    P(h,t) = \int_0^\infty \frac{dV_0}{2\pi V_0} \, e^{-h^2/(2 V_0)}\, \frac{1}{t^2}\, F_0\left(\frac{V_0}{t^2}\right)= \frac{1}{4t}\, F\left(\frac{h}{4t}\right),
    \label{pxt}
\end{equation}
where $F(y)$ is given explicitly by~\cite{SBS2021, SBS2022} (also see~\cite{SGMSS2020} for an alternative expression),  
\begin{equation}
    F(y)=\frac{1}{\sqrt{2}\, \pi^{3/2}} \Gamma\left(\frac{1}{4} + i y\right)\, \Gamma\left(\frac{1}{4} - i y\right),
    \label{eq:Fy}
\end{equation}
where $\Gamma(z)$ is the gamma function.
The scaling function $F(y)$ is symmetric in $y$ with an exponential tail 
\begin{equation}
    F(y) \sim \sqrt{\frac{2}{\pi\, |y|}}\, e^{-\pi |y|}\quad \text{as} ~~ y\to \pm \infty~.
\end{equation}
Thus in this model, 
 while the typical $h$ grows ballistically with $t$, the scaling function of the position distribution has an anomalous exponential tail.

The diffusing diffusivity model with the choice  $D(t)=B^2(t)$ has so far been studied only for a single particle as discussed above. This corresponds to the interaction-free Rouse chain, i.e., for $\kappa=0$ in \eref{eq:Rouse}. It is then natural to ask what happens to the position distribution of a tagged monomer in a Rouse chain in the presence of interaction, i.e., $\kappa >0$ and the choice of   $D(t)=B^2(t)$.  In this paper, we address this question and compute the full height distribution exactly at late times $t\gg 1/\kappa$. This late time behavior of the Rouse model for $D(t)=B^2(t)$ is essentially captured by the continuum EW equation \eref{eq:EW} with $D(t)=B^2(t)$. By performing an exact computation on this variant of the EW model, where $D(t)=B^2(t)$, we show that the scaled height distribution at all times has a nontrivial shape with an exponential tail. This result, valid for the EW model at all times, provides the late time ($t\gg 1/\kappa$) behavior of the scaled position distribution of a tagged monomer in the Rouse chain. 
We will see that  the main technical challenge in this computation involves the study of a particular functional of a Brownian motion that requires a nontrivial modification of the standard Feynman-Kac formalism~\cite{majumdar2005brownian}.  To the best of our knowledge, our result presents the first exact solution of the position distribution of a tagged particle in an interacting many-body system driven by a noise with a time-dependent stochastic diffusivity.

The rest of the paper is organized as follows. In section (\ref{s:model}) we give the definition of the Rouse chain model studied here and recall its continuum formulation as the EW equation. We compute the height variance $V(t)$ for an arbitrary time dependent diffusion constant, this is a non-local function of $D(t)$ and depends on its full history.  The full probability distribution function of the height $h(t)$ at a given point is then derived in terms of the probability density function for $V(t)$. In section (\ref{FK}) the Laplace transform of the probability distribution function of $V(t)$ for the case $D(t)=B^2(t)$ is computed by adapting suitably the backward Feynman-Kac approach. Then, using this result, in section (\ref{pdf}) we show how one can analytically derive the behavior of the probability density function $p(h,t)$ of $h(t)$ for small and large values of $h$.
In \sref{gen--lin-int}, we generalize the computation of the height distribution to a class of linear interface models parametrized by a dynamical exponent $z>1$, with $z=2$ corresponding to the EW case. 
In section (\ref{conc}) we conclude and discuss some possible extensions of our study.

\section{The model}
\label{s:model}

We consider the Rouse dynamics in one dimension, 
\begin{equation}
    \frac{dh_n}{dt} = \kappa\,  [h_{n+1}(t) + h_{n-1}(t) - 2 h_n(t))] + \sqrt{2D(t)}\, \eta_n(t). 
    \label{eq:Rouse1}
\end{equation}
where $D(t)=B^2(t)$ with $B(t)$ representing a Brownian motion in \eref{eq:bm}.  Our main interest is to compute the position distribution of the $n$-th monomer at late times $t\gg 1/\kappa$. 
While this computation can be done, in principle, for the discrete Rouse chain at all times $t$, to extract the late time ($t\gg 1/\kappa$) scaling behavior of the position distribution, it is convenient to consider the continuum EW version
\begin{equation}
    \frac{\partial h}{\partial t} = \Gamma\,  \frac{\partial^2 h}{\partial x^2}  + \sqrt{2 D(t)}\,\eta(x,t)\,,
    \label{eq:EW1}
\end{equation}
with $D(t)=B^2(t)$ and 
$\eta(x,t)$ is a Gaussian white noise with the correlation $\langle \eta(x,t)\eta(x',t')\rangle =\delta(x-x')\delta(t-t')$.  Note that $\Gamma \propto \kappa$ represents the strength of the harmonic interaction. 

Our goal now is to compute the height distribution at a fixed point $x$ at time $t$. 
For simplicity, we start from the flat initial condition $h(x,0)=0$. 
Due to the fact that the noise is Gaussian noise and the EW equation  \eref{eq:EW1} is linear we have that  the height fluctuations are Gaussian for a given history of  $\{D(\tau); 0 <\tau<t\}$  having probability density function
\begin{equation}
    p(h,t|\{D(\tau\})= \frac{1}{\sqrt{2 \pi V(t) }}\, \exp\left( - \frac{h^2}{2 V(t) }\right),
    \label{dist-h}
\end{equation}
where $V(t)=\langle h^2(x,t)\rangle$ denotes the variance for a given realization of the history  $\{D(\tau); 0 <\tau<t\}$.   

To compute this history-dependent  variance, it is convenient to consider the Fourier transform $\tilde{h}(k,t)=\int_{-\infty}^\infty h(x,t) e^{ikx} \, dx$, which  satisfies the equation
\begin{equation}
      \frac{\partial \tilde{h}}{\partial t} = -\Gamma k^2 \tilde{h}  + \sqrt{2 D(t)}\,\tilde{\eta}(k,t),  
\end{equation}
where the noise has the correlation $\langle \tilde{\eta}(k,t)\tilde{\eta}(k',t')\rangle = 2\pi \delta(k+k') \delta(t-t')$. 
The solution of the above equation is given by
\begin{equation}
    \tilde{h}(k,t)= \int_0^t e^{-\Gamma k^2 (t-t')} \sqrt{2 D(t')}\, \tilde{\eta}(k,t').
\end{equation}
It is straightforward to show that 
\begin{equation}
\langle \tilde{h}(k,t) \tilde{h}(k',t) \rangle = 2\pi \delta(k+k') \int_0^t dt'
\, e^{-2 \Gamma k^2 (t-t')} \, 2 D(t'). 
\label{eq:tildeh}
\end{equation}
Consequently,  it follows that
\begin{equation}
    \langle h^2(x,t)\rangle = 
    \int_{-\infty}^\infty \, \int_{-\infty}^\infty\, \frac{dk}{2\pi}\, \frac{dk'}{2\pi}
    \langle \tilde{h}(k,t) \tilde{h}(k',t) \rangle e^{-i (k+k') x}.
    \label{variance1}
\end{equation}
Substituting \eref{eq:tildeh} and carrying out the integrals over $k$ and $k'$, we get
\begin{equation}
   V(t)= \langle h^2(x,t)\rangle = 
    \frac{1}{\sqrt{2\pi \Gamma}}\int_0^t 
    \frac{D(t')}{\sqrt{t-t'}}\, dt'.
    \label{variance2}
\end{equation}

This result is general and holds for arbitrary stochastic diffusivity $D(t)$. For example, for a constant $D(t)=D$, one recovers from \eref{variance2} the result stated in \eref{eq:Vt}.
For the choice $D(t)=B^2(t)$, on which we focus in the rest of the paper, one gets from \eref{variance2}
\begin{equation}
    V(t)=\langle h^2(x,t)\rangle = 
    \frac{1}{\sqrt{2\pi \Gamma}}\int_0^t 
    \frac{B^2(t')}{\sqrt{t-t'}}\, dt',
    \label{variance3}
\end{equation}
where we recall that $B(t)$ is a Brownian motion defined in \eref{eq:bm}. One notices further that the Brownian motion is self-similar, i.e., $B(tu)\equiv\sqrt{t} B(u)$, in the sense that both sides of $\equiv$ have the same statistical distribution. Substituting $t'=tu$ with $u\in[0,1]$ in \eref{variance3}, we then get 
\begin{equation}
    V(t) =\langle h^2(x,t)\rangle = \frac{ t^{3/2}}{\sqrt{2\pi \Gamma}}\int_0^1 
    \frac{B^2(u)}{\sqrt{1-u}}\, du = \frac{ t^{3/2}}{\sqrt{2\pi \Gamma}} \, V.
    \label{variance4}
\end{equation}
where $V$ is a random variable independent of $t$, and is given by  
\begin{equation}
    V = \int_0^1 
    \frac{B^2(u)}{\sqrt{1-u}}\, du, \quad\text{with~~}  B(0)=0.
    \label{BF-V}
\end{equation}
Let us denote by $Q(V)$ the probability distribution of $V$.
Substituting \eref{variance4} in \eref{dist-h} and averaging over the random variable $V$ distributed by $Q(V)$, one gets
an exact scaling form for $p(h,t)$ valid for all $t$, 
\begin{equation}
    p(h,t) = \frac{(2\pi\Gamma)^{1/4} }{ t^{3/4}} G\left( \frac{(2\pi\Gamma)^{1/4} h}{ t^{3/4}}\right),
\end{equation}
where the scaling function $G(H)$, with $H=(2\pi\Gamma)^{1/4} h/ t^{3/4}$ denoting the scaled height, is given by
\begin{equation}
G(H)= \int_0^\infty \frac{1}{\sqrt{2\pi V}}\, \exp\left(-\frac{H^2}{2V}\right)\, Q(V)\, dV.
\label{G(H)}
\end{equation}
Thus to compute this scaling function $G(H)$, we just need to compute the distribution $Q(V)$ of the random variable $V$ in \eref{BF-V}. This can be done exactly by adapting the Feynman-Kac formalism as we show in the next section.

\section{Exact computation of \texorpdfstring{$Q(V)$}{Q(V)}}
\label{FK}
In this section, we compute the distribution $Q(V)$ of the random variable $V$ defined in \eref{BF-V}. This random variable $V$ is actually a functional of the Brownian motion over the scaled time $u\in [0,1]$. Normally, Brownian functionals over a fixed time interval of the type, $\int_0^1 Y\left(B(u)\right)\, du$ where $Y(z)$ is an arbitrary function, can be most conveniently computed using a backward Feynman-Kac formalism, where the initial position of the Brownian motion is considered as a variable~\cite{majumdar2005brownian}. However, in our case, due to the presence of the explicit time dependent factor $1/\sqrt{1-u}$ in the integrand of $V$ in \eref{BF-V}, the standard backward Feynman-Kac approach cannot be used directly. One needs to first adapt this approach to take into account this additional time dependent factor  and we adapt  the method  proposed in \cite{Boyer_2011} to this end.

To proceed, we first define the object
\begin{equation}
\psi_p(x,u) = \left\langle
\exp\left(-p \int_u^1
\frac{B^2(u')}{\sqrt{1-u'}}\, du'\right)
\right\rangle\quad\text{where}~B(u)=x.
\label{eq:psi1}
\end{equation}
We will see below that one can write down an explicit partial differential equation for $\psi_p(x,u)$ and solve it. From this solution, setting $u=0$ and $x=0$, one can compute 
\begin{equation}
    \psi_p(0,0) =  \left\langle
\exp\left(-p \int_0^1
\frac{B^2(u')}{\sqrt{1-u'}}\, du'\right)\right\rangle\quad\text{with}~~ B(0)=0. 
\end{equation}
However, this is just the Laplace transform of $Q(V)$ since 
\begin{equation}
    \psi_p(0,0) = \langle e^{-pV} \rangle = \int_0^\infty e^{-p V} \, Q(V)\, dV. 
    \label{psi00}
\end{equation}
Inverting this Laplace transform $\psi_p(0,0)$ with respect to $p$, one then obtains $Q(V)$.

\begin{figure}
    \centering
    \includegraphics[width=0.5\linewidth]{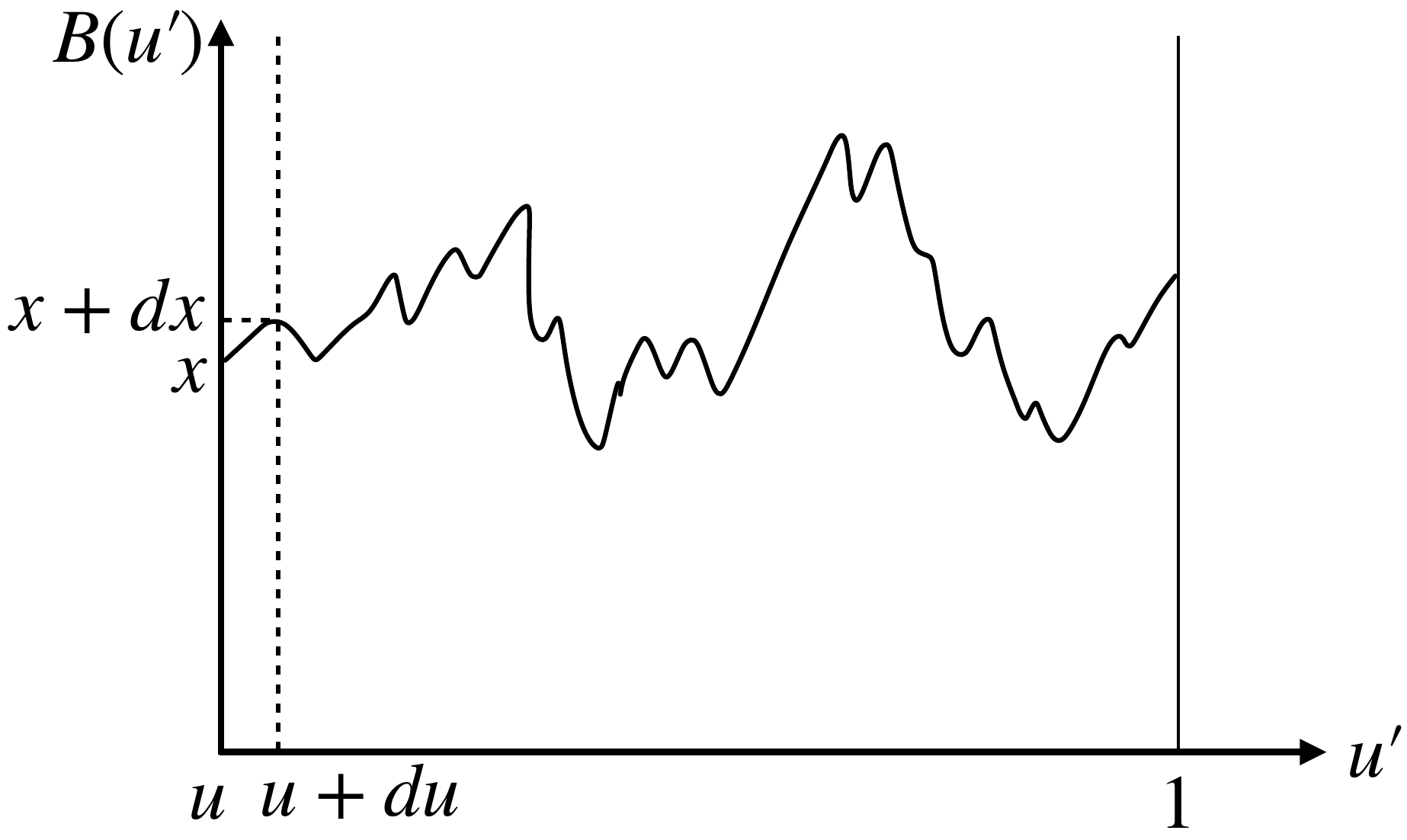}
    \caption{A schematic trajectory of a Brownian motion $B(u')$ propagating from time $u'=u$ to time $u'=1$, starting at $B(u)=x$. In the first infinitesimal time step $du$, the Brownian motion moves from $x$ to $x+dx$.}
    \label{fig:bmtraj}
\end{figure}

To derive the partial differential equation for $\psi_p(x,u)$, defined in \eref{eq:psi1}, we proceed as follows. Here, it is convenient to look at \fref{fig:bmtraj}, where a typical trajectory of a Brownian motion $B(u')$ propagating from $u'=u$ to $u'=1$, starting at $B(u)=x$. 
As shown in this figure, we first split the scaled time interval 
$[u,1]$ into two pieces: $[u,u+du]$ and $[u+du,1]$.  In this first interval of duration $du$, the Brownian motion jumps from the initial position $B(u)=x$ to $B(u+du)=x+dx$ as shown in \fref{fig:bmtraj}. Starting from this new initial position $B(u+du)=x+dx$, the Brownian motion then propagates over the second interval $[u+du, 1]$. We also split the integral into two parts
\begin{equation}
    \int_u^1
\frac{B^2(u')}{\sqrt{1-u'}}\, du' = \int_u^{u+du}
\frac{B^2(u')}{\sqrt{1-u'}}\, du'+ \int_{u+du}^1
\frac{B^2(u')}{\sqrt{1-u'}}\, du'.
\label{eq:bu1}
\end{equation}
To leading order in $du$ as $du\to 0$, the first integral on the right hand side can be approximated as 
\begin{equation}
    \int_u^{u+du}
\frac{B^2(u')}{\sqrt{1-u'}}\, du' = \frac{B^2(u)}{\sqrt{1-u}}\, du + O(du^2).
\label{eq:bu2}
\end{equation}
Substituting \eref{eq:bu1} in \eref{eq:psi1} and using \eref{eq:bu2} and $B(u)=x$, one then gets
\begin{equation}
\psi_p(x,u) = \left(1-\frac{px^2 du}{\sqrt{1-u}}+ O(du^2)\right) \bigl\langle \psi_p(x+dx,u+du) \bigr\rangle_{dx}
\label{eq:psi2}
\end{equation}
where the average is with respect to the initial jump increment $dx$, with $\langle dx\rangle =0$ and $\langle dx^2 \rangle = 2 du$, which follows from the definition of the Brownian motion in  \eref{eq:bm}. Expanding $\psi(x+dx,u+du)$ in Taylor series about $(x,u)$ and taking the limit $du\to 0$, we get the desired  partial differential equation
\begin{equation}
    -\frac{\partial \psi_p (x,u)}{\partial u} = \frac{\partial^2\psi_p (x,u)}{\partial x^2} - \frac{px^2}{\sqrt{1-u}}\psi_p(x,u), \quad\text{valid in~~} u\in [0,1].
    \label{eq:psi3}
\end{equation}
In addition, the solution must satisfy the terminal condition $\psi_p(x,u=1)=1$, which follows from the definition of $\psi_p(x,u)$ in \eref{eq:psi1}.

To solve \eref{eq:psi3}, 
it is convenient to make a change of variable $w=1-u$, which yields
\begin{equation}
    \frac{\partial \phi_p (x,w)}{\partial w} = \frac{\partial^2\phi_p (x,w)}{\partial x^2} - \frac{px^2}{\sqrt{w}}\phi_p(x,w)\quad\text{where~~} \phi_p(x,w)= \psi_p(x,1-w).
    \label{Feynman-Kac}
\end{equation}
This equation is valid for $w\in [0,1]$ with the terminal condition $\psi_p(x,u=1)=1$ translating into the initial condition $\phi_p(x,w=0)=1$.

To solve \eref{Feynman-Kac}, we use the ansatz, 
\begin{equation}
    \phi_p(x,w)= f(w) \exp\left(-\frac{1}{2}g(w) x^2\right)
    \label{ans1}
\end{equation}
where $f(w)$ and $g(w)$ are yet to be determined. Substituting this ansatz \eref{ans1} in \eref{Feynman-Kac}
and matching the coefficients of $x^0$ and $x^2$ we get
\begin{equation}
    \frac{f'(w)}{f(w)} = -g(w) ~~\text{and}~~ -\frac{1}{2} g'(w) =  g^2(w) - \frac{p}{\sqrt{w}}\,. 
    \label{fgeq}
\end{equation}
The condition $\phi_0(x,0)=1$ in \eref{ans1} dictates that we must have $f(0)=1$ and $g(0)=0$. Solving the first equation in \eref{fgeq} using $f(0)=1$, one expresses $f(w)$ in terms of 
 $g(w)$ as 
\begin{equation}
    f(w) = \exp\left(-\int_0^w g(w')\, dw'\right).
    \label{f-g-eq}
\end{equation}

To solve the Riccati equation for $g(w)$ in \eref{fgeq}, we make a Hopf-Cole transformation  $g(w) =  b \, s'(w)/s(w)$ which gives a nonlinear differential equation for $s(w)$
\begin{equation}
    \frac{s''(w}{s(w)} + (2b-1) \left(\frac{s'(w)}{s(w)}\right)^2 - \frac{2}{b} \frac{p}{\sqrt{w}}=0.
    \label{hc}
\end{equation}
Choosing $b=1/2$ reduces this nonlinear equation to a linear second order differential equation for $s(w)$

\begin{equation}
    s''(w) -\frac{4p}{\sqrt{w}}\, s(w) =0, 
    \label{h(w)-eq}
\end{equation}
whose general solution can be obtained exactly as  
\begin{equation}
    s(w)= c_1 \, \sqrt{w} \,I_{2/3}\left(\frac{8}{3} \sqrt{p} \,w^{3/4}\right) + c_2\, \sqrt{w} \,I_{-2/3}\left(\frac{8}{3} \sqrt{p} \,w^{3/4}\right)
    \label{s-sols}
\end{equation}
 where $I_\nu(z)$ is the modified Bessel function of the first kind, and $c_1$ and $c_2$ are two arbitrary constants to be determined shortly.

 Having obtained the general solution of $s(w)$, one can then express the two unknown functions $f(w)$ and $g(w)$  
 in terms of $s(w)$ as
\begin{equation}
    f(w) = \sqrt{\frac{s(0)}{s(w)}}  ~~\text{and}~~ g(w) = \frac{s'(w)}{2 s(w)}\, .
    \label{g-f}
\end{equation}
Next, the boundary condition $g(0)=0$ implies that $s'(0)=0$. Expanding the general solution for $s(w)$ in \eref{s-sols} as $w\to 0$, one gets 
\begin{equation}
    s(w)= c_1\, \left[ \frac{2^{4/3} \, p^{1/3}}{3^{2/3}\, \Gamma(5/3)}\, w + O(w^{5/2})\right] +c_2 \, \left[ \frac{3^{2/3}}{2^{4/3} \,p^{1/3}\, \Gamma(1/3)} + O(w^{3/2})\right]\,.
    \label{sw-sm}
\end{equation}
Hence the derivative at $w=0$ is $s'(0)= c_1 \,  \frac{2^{4/3} \, p^{1/3}}{3^{2/3}\, \Gamma(5/3)}$. The condition $s'(0)=0$ then implies $c_1=0$. The remaining unknown constant $c_2$ is not needed since it cancels out in the expression of $g(w)=s'(w)/(2 s(w))$.  Hence, we obtain
\begin{equation}
    \frac{s(w)}{s(0)}= \frac{2^{4/3}  \Gamma (1/3)}{3^{2/3}}\, p^{1/3}\sqrt{w} \,I_{-2/3}\left(\frac{8}{3} \sqrt{p} \,w^{3/4}\right)\,.
    \label{eq:ss}
\end{equation}
Consequently from \eref{g-f} we get 
\begin{equation}
\label{fw-f}
    f(w) = \sqrt{\frac{s(0)}{s(w)}} = \frac{3^{1/3}}{2^{2/3}  \sqrt{\Gamma (1/3)}}\, \frac{1}{\sqrt{ p^{1/3}\sqrt{w} \,I_{-2/3}\left(\frac{8}{3} \sqrt{p} \,w^{3/4}\right)}}\, .
\end{equation}
Similarly, one can obtain an explicit expression for $g(w)=s'(w)/(2 s(w))$, which however, is rather long. Fortunately, for our purpose, we do not need $g(w)$, and hence, we do not present this complicated expression here. 
Substituting these exact $f(w)$ and $g(w)$ in \eref{ans1} provides us the exact solution for $\phi_p(x,w)$ and consequently for $\psi_p(x,u)=\phi_p(x,1-u)$. Finally, setting $x=0$ and $u=0$, and using \eref{psi00} we get
 the explicit form for  Laplace transform of $Q(V)$ as
\begin{equation}
  \int_0^\infty e^{-pV} Q(V) \, dV = \psi(0,0) = f(1) = \frac{3^{1/3}}{2^{2/3}  \sqrt{\Gamma (1/3)}}\, \frac{1}{\sqrt{ p^{1/3} \,I_{-2/3}\left(\frac{8}{3} \sqrt{p} \right)}} \, .
  \label{LT-Q(V)}
\end{equation}

We use this result in the next section to compute the distribution $G(H)$ of the scaled height using \eref{G(H)}.

\section{Distribution of the scaled height \texorpdfstring{$H$}{H}}
\label{pdf}

It is convenient to rewrite \eref{G(H)} using the Fourier transform of Gaussian as,
\begin{equation}
G(H) = \int_{-\infty}^\infty\frac{dk}{2\pi} e^{ikH} \left[\int_0^\infty e^{-\frac{1}{2} k^2 V} Q(V)\, dV\right].
\label{eq:Gzx}
\end{equation}
We can easily identify the  integral within the square-bracket with the Laplace transform \eref{LT-Q(V)} with $p=k^2/2$. Therefore, 
\begin{equation}
G(H) = \frac{3^{1/3}}{\sqrt{2\Gamma (1/3)}}\int_{-\infty}^\infty\frac{dk}{2\pi}  \frac{e^{ikH}}{\sqrt{ k^{2/3} \,I_{-2/3}\left(\frac{8k}{3\sqrt{2}}  \right)}} .
\label{P(H)-FT}
\end{equation}
Note that the function $ \mathcal{F} (k) =k^{2/3} \,I_{-2/3}\left(\frac{8k}{3\sqrt{2}}  \right)$ is symmetric around $k=0$. Consequently, $G(H)$ is a symmetric function of $H$ (as also evident from \eref{G(H)}). The expression for $G(H)$ in \eref{P(H)-FT} is exact for all $H$. Unfortunately, it is hard to perform the integral to obtain an explicit expression for $G(H)$. However, once can easily extract the asymptotic behaviors of $G(H)$ as shown below.

\begin{figure}
    \centering
    \includegraphics[width=6cm]{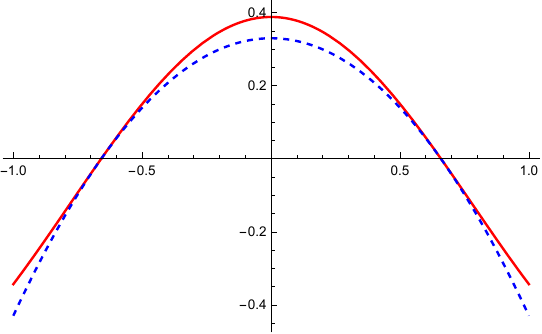}
    \caption{The red solid line plots the function $\mathcal{F} (iq)$ as a function of $q$ whereas the blue dashed line plots the approximation $b (-q^2+a^2)$. Both the solid and the dashed lines are in excellent agreement near $q=\pm a$, with $a = 0.6592248\dots$. }
    \label{fig:bessel-near-singularity}
\end{figure}

\begin{figure}
    \centering
    \includegraphics[width=4in]{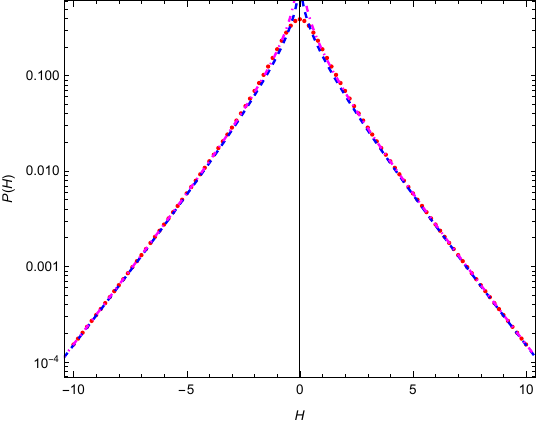}
    \caption{The red points are obtained from the exact numerical integration of \eref{P(H)-FT}. The blue dashed and magenta dot-dashed lines represent the two forms given the last line of \eref{P(H)-largeH} respectively.}
    \label{fig:pH-tail}
\end{figure}

To extract the large $H$ asymptotic behavior of $G(H)$, we need to first locate the singular points of the integrand in the complex $k$ plane. Since the function $G(H)$ is symmetric, we can focus only on large $H>0$. It turns out that the function $ \mathcal{F} (k) =k^{2/3} \,I_{-2/3}\left(\frac{8k}{3\sqrt{2}}  \right)$ has zeros on the imaginary axis, which correspond to having square-root branch points of the integrand when extented to the complex-$k$ plane. For $H>0$, one can deform the integration contour 
in upper-half complex-$k$ plane, passing around these branch points. 
The leading asymptotic behavior of $G(H)$ for large $H$ emerges from the contribution to the integral from 
the closest branch-point singularity at $k=ia$  where $a=0.6592248\dots$ (as can be checked in Mathematica). To extract this leading behavior, we consider the function $\mathcal{F}(k)$ in the vicinity of $k=ia$. In this neighborhood, one finds $\mathcal{F}(k) \approx b (k^2+a^2)$ where  $b =\lim_{k\to ia} \mathcal{F}/(k^2+a^2) = 0.7592287\dots$ [for a numerical verification of this fact, see \fref{fig:bessel-near-singularity}].  
Substituting this approximate $\mathcal{F}(k)$ in 
\eref{P(H)-FT}, we get as $H\to\infty$ 
\begin{align}
G(H) &\approx \frac{3^{1/3}}{\sqrt{2 b\Gamma (1/3)}}
\int_{-\infty}^\infty\frac{dk}{2\pi}  \frac{e^{ikH}}{\sqrt{k^2+a^2}} 
\cr 
&=\frac{3^{1/3}}{\sqrt{2 b\Gamma (1/3)}} \frac{1}{\pi} K_0(a|H|)
\sim \frac{3^{1/3}}{\sqrt{2 b\Gamma (1/3)}} \frac{e^{-a|H|}}{\sqrt{2\pi a |H|}}.
\label{P(H)-largeH}
\end{align}
where $K_0(z)$ is the modified bessel function of the second kind and we recall that $a=0.6592248\dots$ and $b = 0.7592287\dots$. In \fref{fig:pH-tail}, we compare this asymptotic behavior with the exact $G(H)$ obtained by numerically integrating \eref{P(H)-FT}, finding excellent agreement. Thus for large $H$, the scaled height distribution has a non-Gaussian tail in \eref{P(H)-largeH}.

The small $H$ behaviour of $G(H)$ can be easily obtained by expanding $\exp(ikH)$ in \eref{P(H)-FT}, in the power series in $kH$ and performing the integral over $k$  for each term. This yields
\begin{equation}
    G(H) = \sum_{n=0}^\infty
    \frac{(-1)^n d_{n}}{(2n)!} \, H^{2n},
    \label{p(H)-sm}
\end{equation}
where
\begin{equation}
    d_n= \frac{3^{1/3}}{\sqrt{2\Gamma (1/3)}}\int_{-\infty}^\infty\frac{dk}{2\pi}  \frac{k^{2n}}{\sqrt{ k^{2/3} \,I_{-2/3}\left(\frac{8k}{3\sqrt{2}}  \right)}}. 
\end{equation}

\begin{figure}
    \centering
    
\includegraphics[width=4in]{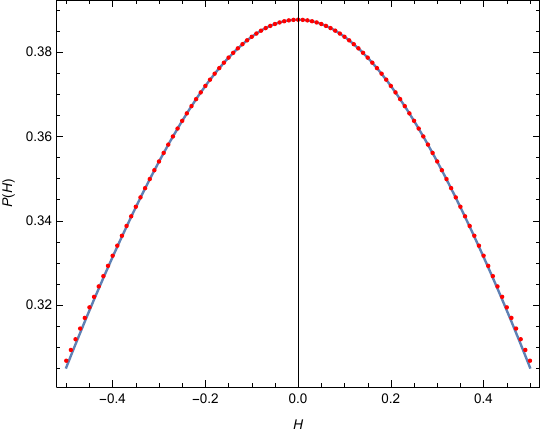}
    \caption{The points are obtained from exact numerical integration of \eref{P(H)-FT}, while the solid line plots \eref{p(H)-sm} keeping upto $O(H^6)$ terms, with the coefficients $d_n$ obtained numerically.}
    \label{fig:pH-small}
\end{figure}

The numerical value of the first few coefficients are given by $d_0=0.387\dots$, $d_1=0.819\dots$, $d_2=10.483\dots$, and $d_3=342.279\dots$. \Fref{fig:pH-small} compare the above small $H$ behavior with the exact distribution obtained numerically from \eref{P(H)-FT}.

Our main result can then be summarized as follows. 
In the presence of a nonzero interaction strength $\Gamma$, the scaled height distribution (equivalently,  the position distribution of a tagged monomer) with a stochastic diffusivity modeled by the square of a Brownian motion, has the scaling behavior 
\begin{equation}
    p(h,t) = \frac{(2\pi\Gamma)^{1/4} }{ t^{3/4}} G\left( \frac{(2\pi\Gamma)^{1/4} h}{ t^{3/4}}\right),
\end{equation}
where the scaling function $G(H)$, symmetric in $H$,  is given exactly by \eref{P(H)-FT} and has the asymptotic behaviors
\begin{equation}
    G(H) \approx \begin{cases}
        \frac{3^{1/3}}{\sqrt{2 b\Gamma (1/3)}} \frac{e^{-a|H|}}{\sqrt{2\pi a |H|}} &\quad\text{as~~} H\to \pm \infty\, ,\\[5mm]
        d_0+ \frac{d_2}{24} H^2 &\quad\text{as~~} H\to 0\, .
    \end{cases}
    \label{GH-asym}
\end{equation}
Thus the scaled height distribution in this model exhibits a nontrivial shape with an exponentially decaying tail. 

\section{Generalization to other linear interface models}
\label{gen--lin-int}

In this section, we generalize the computation of the height distribution with $D(t)=B^2(t)$ for other one-dimensional linear interface models, such as
\begin{equation}
    \frac{\partial h}{\partial t} = -\Gamma\, \left(- \partial_x^2 h\right)^{z/2} + \sqrt{2 \, D(t)}
\, \eta(x,t)\, ,
\label{eq:gen-in}
\end{equation}
where the dynamical exponent $z$ characterizes the interface dynamics and $\eta(x,t)$ is a Gaussian white noise as before. For $z=2$, one recovers the EW equation~\eref{eq:EW}. With a constant $D(t)$, such interface models for other values of $z$ have been studied extensively in the literature~\cite{ krug1997origins, bray2013persistence, majumdar2001spatial, dean2021position}.
The case $z=4$ corresponds to the Mullins-Herring equation for surface growth~\cite{mullins1957theory, herring1950effect} and is also related to semi-flexible polymer chain~\cite{dean2021position}. The other values of $z$ also have interesting applications, see e.g.,~\cite{bray2013persistence, majumdar2001spatial, dean2021position}. Here, we compute the height distribution in an infinite one-dimensional system with $D(t)=B^2(t)$. 

The calculation for general $z$ proceeds more or less the same way as the EW case ($z=2$) presented in the previous section. Here, we briefly outline the main steps for general $z>1$.  Taking Fourier transform of the height function in \eref{eq:gen-in}, we find the correlator
\begin{equation}
    \langle \tilde{h}(k,t)\, \tilde{h}(k',t)\rangle = 2\pi\,  \delta(k+k')
    \, \int_0^t e^{-2 \Gamma |k|^z (t-t')}\, 2 D(t')\, dt'\, .
    \label{eq:tilhg}
\end{equation}
For $z=2$, it reduces to~\eref{eq:tildeh}. Consequently, the variance is given by 
\begin{equation}
    V(t)=\langle h^2(x,t)\rangle = 
    \int_{-\infty}^\infty \, \int_{-\infty}^\infty\, \frac{dk}{2\pi}\, \frac{dk'}{2\pi}
    \langle \tilde{h}(k,t) \tilde{h}(k',t) \rangle e^{-i (k+k') x}.
    \label{gvariance1}
\end{equation}
Substituting \eref{eq:tilhg} and carrying out the integrals over $k$ and $k'$, we get
\begin{equation}
   V(t)= \langle h^2(x,t)\rangle = A_z\, 
   \int_0^t 
    \frac{D(t')}{(t-t')^{1/z}}\, dt'\quad\text{where}\quad A_z=\frac{2^{1-1/z}\, \Gamma(1/z)}{\pi \, z\, \Gamma^{1/z}}.
    \label{gvariance2}
\end{equation}
where $\Gamma(x)$ is the standard Gamma function and should not be confused with the coupling coefficient $\Gamma$. For $z=2$, it reduces to~\eref{variance2}. Setting   $D(t)=B^2(t)$, where $B(t)$ is given in~\eref{eq:bm}. Following exactly the same steps as in the $z=2$ case, one gets 
\begin{equation}
    V(t)=\langle h^2(x,t)\rangle = A_z\, t^{2-1/z}\, 
    \int_0^1 
    \frac{B^2(u)\,  du}{(1-u)^{1/z}}\, \quad\text{with~~}  B(0)=0\, .
    \label{gvariance3}
\end{equation}
Here we have assumed that $z>1$ such that the integral converges. For $0 < z <1$, one needs to keep a lattice constant (i.e., an ultraviolet cutoff in the $k$ integral) and we will not discuss this case here.

Consequently, the height distribution $p(h,t)$ again can be written in a scaling form
\begin{equation}
    p(h,t) = \frac{1}{\sqrt{A_z}\, t^{1-1/(2z)}}\, G_z\left(\frac{h}{\sqrt{A_z}\, t^{1-1/(2z)}}\right)~,
\end{equation}
where the $z$-dependent scaling function can be written as
\begin{equation}
    G_z(H)=\int_0^\infty \frac{1}{\sqrt{2\pi V}}\, \exp\left(-\frac{H^2}{2V}\right)\, Q_z(V)\, dV~,
\label{gG(H)}
\end{equation}
where $Q_z(V)$ is the PDF of the functional
\begin{equation}
    V=\int_0^1 
    \frac{B^2(u)\,  du}{(1-u)^{1/z}}\, \quad\text{with~~}  B(0)=0\, .
    \label{gF-V}
\end{equation}
The probability distribution of $Q_z(V)$ of the functional $V$ in \eref{gF-V} can again be carried out by following the adapted backward Feynman-Kac approach that we used for the $z=2$ case in~\sref{FK}. We omit the details and state the main results. We find that the Laplace transform of $Q_z(V)$ is given by
\begin{equation}
    \int_0^\infty\, e^{-p\,V}\, Q_z(V)\, dV
=\sqrt{\frac{s_z(0)}{s_z(1)}}~,
\label{eq:gLT}
\end{equation}
where $s_z(w)$ satisfies the differential equation 
\begin{equation}
    s_z''(w) -\frac{4\, p}{w^{1/z}}\, s_z(w)=0\quad\text{for~~} 0\le w\le 1\, ,
    \label{gh(w)-eq}
\end{equation}
with the boundary condition $s_z'(0)=0$. 
For $z=2$, this coincides with \eref{h(w)-eq}. Solving this equation explicitly with the boundary condition $s_z'(0)=0$ gives for $z>1$
\begin{equation}
    \frac{s_z(w)}{s_z(0)}= 2^{-z/(2z-1)}\, \Gamma\left(\frac{z-1}{2z-1}\right)\,
    \left(\frac{4z\, \sqrt{p}}{2z-1}\right)^{z/(2z-1)}
    \sqrt{w} \,I_{-z/(2z-1)}\left(\frac{4z\, \sqrt{p}}{2z-1}\, w^{(2z-1)/(2z)}\right)\,.
    \label{eq:sz}
\end{equation}
For $z=2$, one recovers \eref{eq:ss}. Setting $w=1$ in \eref{eq:sz}, we get from \eref{eq:gLT}
\begin{equation}
    \int_0^\infty\, e^{-p\,V}\, Q_z(V)\, dV = B_z\,\frac{p^{-z/(8z -4)}}
    {\sqrt{
     I_{-z/(2z-1)}\left(\frac{4z\, \sqrt{p}}{2z-1}\right)
     }}
    \,. 
    \label{eq:ggLT}
\end{equation}
where 
\begin{equation}
    B_z= 2^{z/(4z-2)}\, \left[\Gamma\left(\frac{z-1}{2z-1}\right)\right]^{-1/2}\,
    \left(\frac{4z}{2z-1}\right)^{-z/(4z-2)}~.
    \label{eq:Bz}
\end{equation}
For $z=2$, \eref{eq:ggLT} reduces to the earlier result~\eref{fw-f}.

Finally, the scaling function $G_z(H)$ in \eref{gG(H)}, using a Fourier representation of the Gaussian $e^{-H^2/(2V)}/\sqrt{2\pi V} $ and the result in \eref{eq:ggLT},  can be expressed explicitly as
\begin{align}
G_z(H) &= \int_{-\infty}^\infty\frac{dk}{2\pi} e^{ikH} \left[\int_0^\infty e^{-\frac{1}{2} k^2 V} Q_z(V)\, dV\right]\cr
& = B_z \, \int_{-\infty}^\infty\, \frac{dk}{2\pi}\, 
\frac{e^{i k H}}{
\sqrt{
k^{z/(2z-1)}\, I_{-z/(2z-1)}\left(\frac{4z\, k}{2z-1}\right)}
}~,
\label{eq:Gzf}
\end{align}
with $B_z$ given in \eref{eq:Bz}. Once again, it is easy to check that for $z=2$, \eref{eq:Gzf} reduces to \eref{eq:Gzx}. The scaling function $G_z(H)$ is symmetric in $H$. One can derive its asymptotic behavior for general $z$ as in the $z=2$ case, but we do not repeat it here. It is not difficult to show that for large $|H|$, the scaling function $G_z(H)$ has an exponential tail as in $z=2$, but with a $z$-dependent decay exponent. Thus the exponential tail of the height distribution is quite generic.

\section{Conclusion}
\label{conc}

In this paper, we have studied the dynamics of an infinitely long one-dimensional Rouse polymer chain in the presence of a stochastic diffusivity $D(t)=B^2(t)$, where $B(t)$ represents a Brownian motion. Our goal was to compute the position distribution of a tagged monomer at late times $t\gg 1/\kappa$, where $\kappa$ represents the strength of the harmonic interaction between neighboring monomers. This late-time position distribution of a tagged monomer can be derived by studying a simpler continuum version of this dynamics, where the position of the tagged monomer becomes equivalent to the height $h(x,t)$ of a one-dimensional Edwards-Wilkinson interface in the presence of a stochastic diffusivity $D(t)=B^2(t)$. In this paper, we have shown that the height distribution in the latter model can be solved exactly at all times $t$, which then provides the position distribution of a tagged monomer of the Rouse chain at late times. This exact calculation involved computing the distribution of a Brownian functional which required a nontrivial adaptation of the standard backward Feynman-Kac formalism. Our exact calculation shows that the typical height at a given point in space scales as $t^{3/4}$ and the distribution $G(H)$ of the scaled height $H=h/t^{3/4}$ is symmetric and has a nontrivial shape: while it approaches
a nonzero constant quadratically as $H\to 0$, it has a non-Gaussian tail that decays exponentially for large $H$. We then generalized our result for a family of linear interface models parametrized by a dynamical exponent $z>1$ (with $z=2$ corresponding to the Edwards-Wilkinson case). 

Our work can be extended in several future directions. Here we have studied the height distribution of a simple linear interface model in an infinite system in the presence of stochastic diffusivity $D(t)=B^2(t)$. 
It would be interesting to extend this study to a finite system of size $L$. In particular, in the context of the tagged monomer in the Rouse polymer chain of finite size $L$, for which there have been other studies of the Brownian yet non-Gaussian diffusion of the center of mass~\cite{Panja_2010, Nampoothiri2021, Nampoothiri_2022}, it would be interesting to see the effects of diffusing diffusivity.  
Furthermore, it would be interesting to extend our calculation to other forms of $D(t)$, e.g., $D(t)$ may represent the square of an Ornstein-Uhlenbeck process.
It would also be interesting to study the effects of interaction between monomers when they are driven by a switching diffusion process, as in Ref.~\cite{gueneau2025large}, where only a single particle was studied.

While here we focused on the statistics of the interface height at a given point in space, it would be interesting to study the correlations between heights and, more generally, the joint distribution of the heights at different spatial points in the presence of a stochastic diffusivity. Besides,  here we studied simple linear interface models, and it would be of interest to extend this study to nonlinear interface models such as the well-known Kardar-Parisi-Zhang (KPZ) model\cite{kpz} with a stochastic diffusivity $D(t)$. Finally, it would be interesting to study the height distribution in higher dimensions in the presence of a stochastic diffusivity.

\section{Acknowledgements}
 SNM and SS acknowledge the support from the Science and Engineering Research Board (SERB, Government of India), under the VAJRA faculty scheme No. VJR/2017/000110. DSD and SNM acknowledge the ANR Grant No. ANR- 23- CE30-0020-01 EDIPS.

\vskip0.5cm

\section*{References}


\providecommand{\newblock}{}

\end{document}